\documentclass[useAMS,usenatbib,usegraphicx
]{mn2e}
\usepackage{amssymb}
\usepackage{aas_macros}

\newcommand{\Teff}{$T_{eff}$}
\newcommand{\Vt}{$v_{mic}$}
\newcommand{\lgg}{\rm{log}\,$g$}

\newcommand{\Mo}{$M_{\odot}$}

\newcommand{\kms}{km\,s$^{-1}$}
\newcommand{\pz}{PZ~Mon}
\newcommand{\vr}{$v_{\rm rad}$}
\newcommand{\vm}{$v_{mac}$}

\newcommand{\rs}{RS~CVn}

\title[Overabundance of s-process elements in \pz] {Overabundance of
s-process elements in the atmosphere of the active red giant \pz\thanks{Based
on observations collected at the BTA telescope (Special Astrophysical
Observatory, Russia)}}

\author[Yu.V.Pakhomov]{
Yu.~V.~Pakhomov,$^{1}$
 \\
$^1$Institute of Astronomy, Russian Academy of Sciences, Pyatnitskaya 48,
119017, Moscow, Russia\\
}

\begin{document}
\maketitle

\begin{abstract}
Based on high-resolution ($R$=60\,000) spectra taken with the NES spectrograph
(the 6-m BTA telescope, the Special Astrophysical Observatory of the Russian
Academy of Science), we have determined the abundances of 26 elements, from
lithium to europium, in the atmosphere of the active red giant \pz\, which
belongs to the class of \rs\ variable stars, by the method of model stellar
atmospheres. We have taken into account the hyperfine splitting, the isotopic
shift, and the departure from local thermodynamic equilibrium. Analysis of our
data has revealed an overabundance of lithium and neutron-capture elements
compared to normal red giants. For lithium, this is explained by the activity of
the star, while the overabundance of s-elements is presumably similar in nature
to that in moderate barium stars.
\end{abstract}


\section{Introduction}

\pz\ (HD 289114) is an \rs\ variable binary star whose components are
represented by a red giant of spectral type K2III \citep{2015MNRAS.446...56P}
and a red dwarf of spectral type M7V \citep{2015AstL...41..677P}. Previously
\citep{2015MNRAS.446...56P}, we determined the parameters of the stellar
atmosphere for the primary component of \pz\ by various independent methods, its
effective temperature \Teff=4700$\pm$100~K and surface gravity
\lgg=2.8$\pm$0.2, and revealed the signatures of chromospheric activity based on
optical and ultraviolet observations and a corona based on X-ray observations.
The optical variability ($\Delta{}m\approx0.03^m$) of the star is caused by a
nonuniform distribution of temperature spots on its surface: a large fraction of
them is concentrated on the side facing the secondary component, about 24\% of
the surface is covered with spots, while their fraction on the opposite side is
about 20\% \citep{2006A&AT...25..247A}.

The binary system \pz\ is interesting in many aspects. The rotation velocity of
the primary component, \vr=10.5~\kms, exceeds the typical rotation velocities of
red giants, 3--5~\kms. The component mass ratio, $M_2/M_1$ = 0.14\Mo:1.5\Mo =
0.09, is minimal among the known \rs\ giants. At the same time, the binary
system rotates synchronously with a period of 34.13 days, while all such binary
systems similar in parameters are asynchronous \citep{2015AstL...41..677P},
which raises the question about the synchronization mechanism of \pz. Thus, a
comprehensive study of the star \pz\ is topical, including a comparative
analysis of the chemical composition that can set \pz\ apart from the normal red
giants and active \rs\ giants. 

In this paper, we have determined the abundances of 26 elements in the
atmosphere of \pz\ for the first time to perform a comparative analysis of this
active giant with other \rs\ giants and normal red giants of the Galactic thin
disk. These data will subsequently be used to calculate the distribution of
temperature spots on the surface of the star being investigated by the method of
Doppler tomography and to study the influence of the secondary component on the
activity of the primary one.

\section{Observations}

The spectroscopic observations of \pz\ were performed on February 10, 2015, with
the NES spectrograph mounted at the Nasmyth focus of the 6-m BTA telescope at
the Special Astrophysical Observatory of the Russian Academy of Sciences. The
regime of observations with an image slicer was used; the spectral resolution
was $R$=60\,000. Three consecutive half-hour exposures were taken on an e2v
CCD42-90 (4632$\times$2068 pix). We applied the MIDAS software package to
process the spectra. We extracted 54 echelle orders in the spectral range from
3890 to 6980~\AA. The wavelength calibration was based on the spectrum of a
thorium-argon lamp. The preprocessing of the spectra also included the removal
of cosmic-ray particle hits and allowance for the scattered light. An arbitrary
echelle order consists of three separate spectra: the main spectrum and two
spectra from the image slicer, which were processed separately. At the end of
the processing procedure, the signal from the slicer was added to the main
spectrum. Each of the three exposures of the spectrum for \pz\ was also
processed separately, and all spectra were then added with allowance made for
the difference in the Earth's projected rotation velocity. The mean
signal-to-noise ratio in the resulting spectrum was about of 120. The spectrum
was normalized to the continuum level using the blaze function obtained from the
flat field spectrum and using the synthetic spectrum of a star with model
parameters similar to those of \pz. Our primary analysis of the spectrum allowed
us to determine the radial velocity \vr=23.91$\pm$0.16~\kms.

\section{Determining the abundances of chemical elements}

\begin{table}
\caption{Abundances of chemical elements in the atmosphere of \pz: $N$ is the
number of lines used, [El/N] and $\sigma$[El/H] are the relative abundance and
its error, hfs stands for hyperfine splitting, iso stands for isotopic shift,
nLTE means that LTE is abandoned.}
\label{tab:abund}
\centering
\begin{tabular}{|l|c|r|c|l|}
\hline
Element & $N$ & [El/H] & $\sigma$[El/H] & Remark \\
\hline
Li1 &  1 &  0.46 & 0.07 & hfs, iso, nLTE\\   
 C1 &  3 & -0.26 & 0.07 & \\
 N1 &  1 &  0.25 & 0.15 & \\
 O1 &  1 &  0.14 & 0.08 & \\
Na1 &  2 &  0.19 & 0.02 & hfs, nLTE \\
Mg1 &  2 &  0.04 & 0.10 & nLTE \\
Al1 &  1 &  0.00 & 0.03  & hfs, nLTE \\
Si1 &  9 &  0.04 & 0.09 & \\
Ca1 &  6 &  0.05 & 0.08 & \\
Sc1 &  4 &  0.07 & 0.05 & hfs \\
Sc2 &  4 &  0.19 & 0.10 & hfs \\
Ti1 & 26 &  0.07 & 0.08 & \\
Ti2 &  3 &  0.07 & 0.03 & \\
 V1 & 23 &  0.09 & 0.07 & hfs \\
Cr1 & 11 &  0.11 & 0.08 & \\
Mn1 &  5 &  0.13 & 0.11 & hfs \\
Fe1 & 90 &  0.08 & 0.09 & \\
Fe2 &  4 &  0.17 & 0.10 & \\
Co1 & 11 &  0.06 & 0.11 & hfs \\
Ni1 & 14 &  0.04 & 0.08 & \\
Sr1 &  1 &  0.15 & 0.11 & \\
 Y1 &  1 &  0.09 & 0.03 & hfs \\
 Y2 &  2 &  0.18 & 0.06 & hfs \\
Zr1 &  4 & -0.02 & 0.09 & \\
Mo1 &  1 &  0.18 & 0.10 & \\
Ba2 &  3 &  0.58 & 0.10 & hfs, iso, nLTE \\
La2 &  1 &  0.39 & 0.04 & hfs \\
Ce2 &  2 &  0.37 & 0.03 &  \\
Nd2 &  4 &  0.57 & 0.08 &  \\
Eu2 &  1 &  0.22 & 0.08 & hfs, iso \\
\hline
\end{tabular}
\end{table}

\begin{table}
\caption{The errors in the elemental abundances determined by changing the
effective temperature (+100~K), surface gravity (+0.2), and microturbulent
velocity (+0.15~\kms) and the total root-mean-square error}
\label{tab:error}
\centering
\begin{tabular}{|l|r|r|c|c|}
\hline
Ion & \multicolumn{4}{c|}{\small $\Delta$~El/H} \\
\cline{2-5}
        & \small $\Delta$\Teff {\tiny(+100~K)} & \small $\Delta$\lgg
{\tiny(+0.2)} & \small $\Delta$\Vt {\tiny(+0.15~\kms)} & \small total \\
\hline
Li1 &   0.15 &   0.00 &  -0.02 &   0.15 \\
 C1 &  -0.02 &   0.04 &  -0.01 &   0.05 \\
 N1 &   0.12 &   0.06 &   0.00 &   0.13 \\
 O1 &   0.01 &   0.08 &   0.00 &   0.08 \\
Na1 &   0.08 &  -0.02 &  -0.05 &   0.10 \\
Mg1 &   0.03 &  -0.01 &  -0.03 &   0.04 \\
Al1 &   0.07 &   0.00 &  -0.01 &   0.07 \\
Si1 &  -0.05 &   0.04 &  -0.02 &   0.07 \\
Ca1 &   0.10 &  -0.03 &  -0.07 &   0.13 \\
Sc1 &   0.15 &   0.01 &  -0.03 &   0.15 \\
Sc2 &  -0.01 &   0.08 &  -0.05 &   0.09 \\
Ti1 &   0.14 &   0.00 &  -0.05 &   0.15 \\
Ti2 &  -0.02 &   0.08 &  -0.07 &   0.11 \\
 V1 &   0.17 &   0.01 &  -0.07 &   0.18 \\
Cr1 &   0.11 &   0.00 &  -0.04 &   0.12 \\
Mn1 &   0.10 &  -0.00 &  -0.09 &   0.13 \\
Fe1 &   0.04 &   0.02 &  -0.06 &   0.07 \\
Fe2 &  -0.11 &   0.12 &  -0.04 &   0.17 \\
Co1 &   0.04 &   0.04 &  -0.07 &   0.09 \\
Ni1 &   0.03 &   0.03 &  -0.08 &   0.09 \\
Sr1 &   0.19 &   0.00 &  -0.13 &   0.23 \\
 Y1 &   0.18 &   0.00 &  -0.04 &   0.18 \\
 Y2 &  -0.01 &   0.09 &  -0.01 &   0.09 \\
Zr1 &   0.18 &   0.01 &  -0.04 &   0.18 \\
Mo1 &   0.14 &   0.00 &  -0.02 &   0.14 \\
Ba2 &   0.03 &   0.02 &  -0.06 &   0.07 \\
La2 &   0.02 &   0.09 &  -0.02 &   0.09 \\
Ce2 &   0.00 &   0.08 &  -0.01 &   0.08 \\
Nd2 &   0.01 &   0.08 &  -0.03 &   0.09 \\
Eu2 &   0.00 &   0.09 &  -0.01 &   0.09 \\
\hline
\end{tabular}
\end{table}

We analyzed the spectrum by the synthetic spectrum method in the Binmag3 code
(written by O. Kochukhov\footnote{http://www.astro.uu.se/\~{}oleg/binmag.html}).
For our calculations of the theoretical spectra, we used the Synth3 and the
model stellar atmosphere with parameters \Teff=4700$\pm$100~K, \lgg=2.8$\pm$0.2,
and \Vt=1.3~\kms\ computed with the ATLAS9 code \citep{1993KurCD..13.....K}. The
list of atomic and molecular spectral lines for the Synth3 code was retrieved
from the VALD3 database \citep{2015PhyS...90k4005R} using the ``select stellar''
function, which allows one to select only those lines in a given wavelength
range whose intensity exceeds a given threshold value and will be sufficient for
them to be visible in the spectrum. The synthetic spectrum includes 25\,123
lines in the range from 4000 to 7000~\AA\ with an intensity differing from the
continuum intensity by more than 0.01. From this list we selected the lines with
a minimal degree of blending with neighboring lines to determine the abundances
of chemical elements from the condition: 

$$ 
F^{cen} < 0.05 F^{blend} 
$$
where 
$$
F^{blend} = \sum\limits_{i=1}^N
F^{cen}_i e^{-\frac{(\lambda-\lambda_i)^2}{\Delta\lambda^2}}
$$
$$
\Delta\lambda=\sqrt{(\lambda\,v_{rot}\,{\rm sin}\,i/c)^2 + (\lambda/R)^2 }
$$

where $F^{cen}$ is the central depth of the spectral line, $F^{blend}$ is the
estimated depth of the spectrum at the center of the line being investigated
with wavelength $\lambda$ formed by $N$ neighboring lines with wavelengths
$\lambda_i$, $\Delta\lambda$ is the characteristic width of the spectral line,
\vr=10.5~\kms\ is the rotation velocity of \pz\, sin\,$i$=0.92 is the
inclination of the rotation axis of \pz, $c$ is the speed of light, and
$R$=60\,000 is the spectral resolution. The blending was taken into account
around each line at a distance of 5$\Delta\lambda$. From the produced list of
lines we then selected only those that could be measured with a high quality in
the observed spectrum of the star being investigated. The final list includes
294 spectral lines.

 To determine the abundances of chemical elements, we applied the
Levenberg--Marquardt method of fitting the theoretical spectrum to the observed
one. The variation parameters were the elemental abundance log (El/H), the
radial velocity \vr, and the macroturbulent velocity \vm. The quality of the fit
was controlled by the condition under which the difference of the theoretical
and observed spectra was below the noise level. We separately calculated the
profile of a specific spectral line to analyze the degree of influence of the
neighboring lines. All of the measured lines in the spectrum of \pz\ were also
measured in the NOAO solar spectrum \citep{1984sfat.book.....K} for the
subsequent calculation of differential abundances to reduce the systematic
errors in atomic parameters. 

The hyperfine splitting (HFS) effect was taken into account using the hyperfine
structure constants $A$ and $B$ for the atomic levels of odd isotopes of sodium,
aluminum, scandium, vanadium, manganese, cobalt, yttrium, barium, lanthanum, and
europium~\footnote{http://kurucz.harvard.edu/atoms.html}. For this purpose, we
wrote a code that analyzed the input list of spectral lines to calculate the
synthetic spectrum, checked the availability of the necessary data (identified
the lower and upper transition levels, sought for the coefficients $A$ and $B$),
and launched the calculation of the wavelengths and intensities of the HFS
components. As a result, each unsplited line from the input list having data on
its splitting was replaced by a set of lines according to the number of
components with the logarithms of their oscillator strengths ${\rm log}\,gf_0 +
log(I_i) - log(\sum I_i)$, where ${\rm log}\,gf_0$ is the oscillator strength of
the unsplited line, $I_i$ is the relative intensity of the $i$-th component, and
$\sum I_i$ is the sum of the intensities of all splitting components. 

\begin{figure}
\centering
\includegraphics[width=0.49\textwidth,clip]{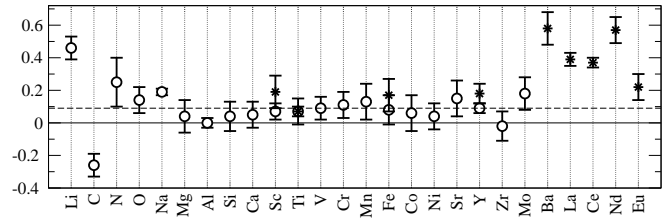}
\caption{Abundances of chemical elements in the atmosphere of \pz. The
circles and asterisks denote the elements whose abundances were determined from
the spectral lines of neutral atoms and ions, respectively. The dashed line
marks the metallicity level determined from iron-peak elements.}
\label{fig:abund}
\end{figure}

\begin{figure*}
\centering
\includegraphics[height=0.65\textheight,clip]{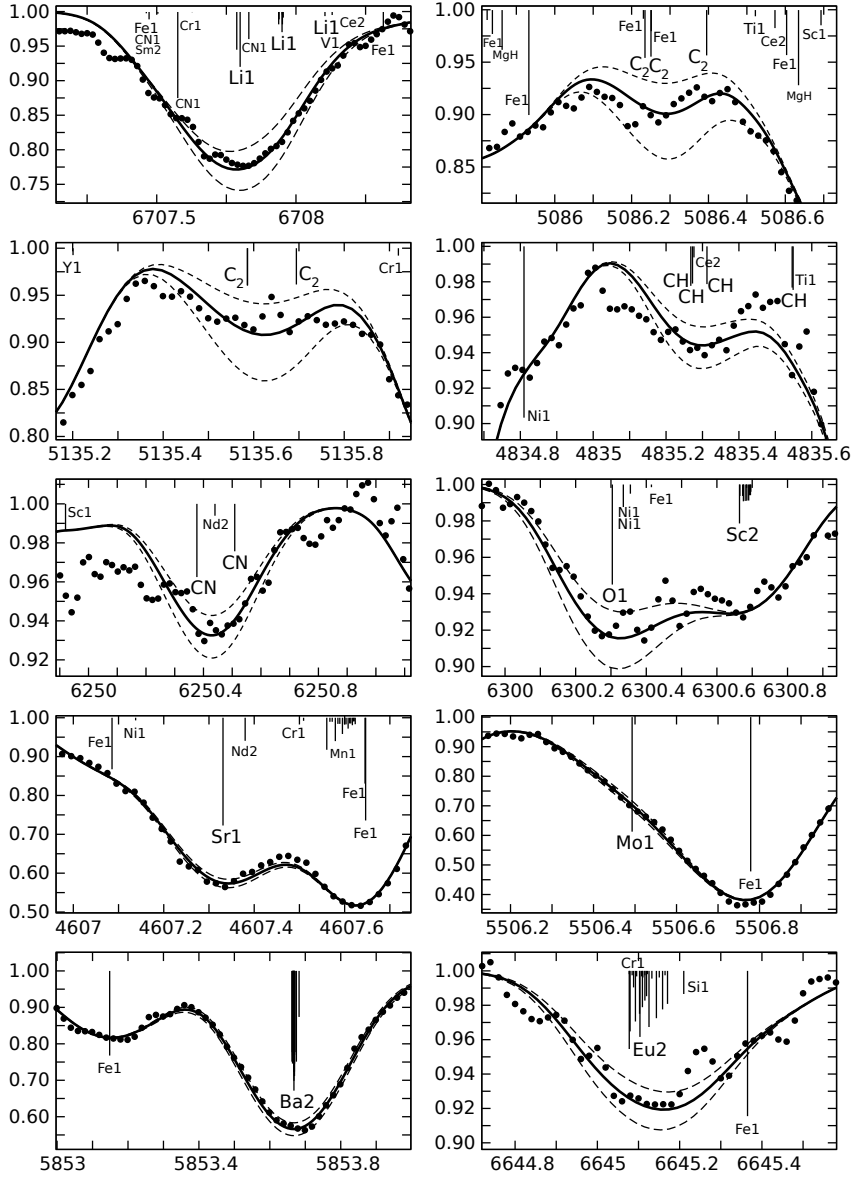}
\caption{Li1, C$_2$ , CH, CN, O1, Sr1, Mo1, Ba2, and Eu2 line profiles. The dots
indicates the observed spectrum, the thick line indicates the best fit synthetic
spectrum, the dotted line indicates the profile when changing the abundance of a
given element by $\pm$0.10, and the vertical lines indicate the relative
contribution to the synthetic spectrum from various lines.}
\label{fig:profile}
\end{figure*}

For lithium, we used the available data on the HFS and isotopic shift by
\citet{1998ApJ...506..405S}. For barium, we performed our calculations
separately for each isotope and then added the values corresponding to the
logarithm of the relative isotope abundance \citep{1998JPCRD..27.1275R} to the
oscillator strengths. We took into account the departure from local
thermodynamic equilibrium (LTE) using non-LTE corrections: from
\cite{2009A&A...503..541L} for lithium, from \cite{2014AstL...40..406A} and
\cite{2007MNRAS.382..553L} for sodium, magnesium, and aluminum, and from
\cite{1996Ap&SS.236..185M} for barium. 

The results are presented in Table~\ref{tab:abund}: the first column gives the
element and its ionization stage from $N$ lines (the second column) of which the
relative abundance [El/H] (the third column) and its error $\sigma$[El/H] (the
fourth column), which also takes into account the accuracy of fitting the line
profiles, were determined; the last column gives the effects that were taken
into account for the lines of the specified element. In the case of nitrogen,
the error in the abundance being determined allows for the error in the carbon
abundance, because nitrogen bonds with carbon into CN molecular lines. These
data are also shown in Fig.~\ref{fig:abund}, where the circles and asterisks
denote the elements whose abundances were determined from the lines of neutral
atoms and ions, respectively.

 To illustrate the reliability of the measurement results,
Fig.~\ref{fig:profile} compares the observed and synthetic profiles for some
spectral lines that are difficult to measure and the profiles when changing the
elemental abundance by $\pm$0.10. The lithium abundance was derived from one
6707~\AA\ line at a distance of 4~\AA\ from the edge of the echelle order. The
line in the spectrum from \cite{2015MNRAS.446...56P} was located slightly
farther, but the signal-to-noise ratio was smaller. Therefore, in this paper, we
averaged the abundances derived by analyzing the previous (-10.31) and present
(-10.46) spectra. When describing the lithium lines, we took into account the
cyanogen molecular lines. We estimated the carbon and nitrogen abundances by CH
(4835~\AA), C$_2$ (5086~\AA\ É 5135~\AA), and CN (6250--6251~\AA) molecular
lines. The first line is located in the region of the spectrum where the
continuum is difficult to draw and gives a low abundance, C/H=-3.80. The C$_2$
lines are very sensitive to the carbon abundance and give similar results,
C/H=-3.70 and -3.64, respectively. The 6300~\AA\ line was the only possible one
for determining the oxygen abundance. This line is blended with two lines of
nickel isotopes ($\lambda_1$=6300.335~\AA, log\,$gf_1=-2.25$ and
$\lambda_2$=6300.355~\AA, log\,$gf_2=-2.67$), a weak iron line, and the scandium
line with HFS affects the line wings. The strontium and molybdenum abundances
were measured from the 4607.33 and 5506.45~\AA\ by taking into account the
possible blends. 

Table~\ref{tab:error} presents the results of our analysis of the errors in the
abundances of chemical elements. The first column gives the ion name, the next
three columns give the changes in elemental abundance when changing the
effective temperature by 100~K, the surface gravity by 0.2~dex, and the
microturbulent velocity by 0.15~\kms, the last column gives the total error
calculated as the root-mean-square error.

\section{Discussion}

Since \rs\ stars have temperature spots on the surface, we estimated their
influence on the observed spectrum of \pz. Our calculations of the theoretical
fluxes (the Synth3 code) show that the stellar continuum intensity in the range
5000--6000~\AA\ for an effective temperature of 4700~K is a factor of 5--6
higher than that for a spot temperature of 3500~K (we took the minimum
temperature of the ATLAS9 models). At the star's filling factor of 0.25 (the
ratio of the spot area to the entire visible disk area), the difference between
the intensities of the surface of a star with and without spots will reach a
factor of 15--18. The presence of such a small number of spots will manifest
itself as a reduction in the continuum level, forming a pseudo-continuum
relative to which the line intensities barely change: the calculated differences
are smaller than the noise of the observed spectrum. If the spot temperature is
even lower, as was pointed out by \cite{2006A&AT...25..247A}, then the influence
of spots on the observed spectrum will be even smaller. The degree of influence
of spots can also be estimated from the presence or absence of strong TiO
molecular bands typical for a stellar atmosphere with a temperature below
4000~K. There are no TiO molecular lines, including those in the strong
6158~\AA\ band, in the spectrum of \pz. Therefore, we analyzed the abundances of
chemical elements without allowance for the spottedness. We also analyzed the
widths of spectral lines depending on their Lande factors. However, no such
dependence was found; therefore, we disregarded the magnetic field in our
calculations.

Because of the cool atmosphere and the fairly high rotation velocity
(10.5~\kms), there are few unblended lines in the spectrum of \pz. For most
rare-earth elements, we still had to use the line profile measurements by taking
into account the blends (Fig.~\ref{fig:profile}). It can be seen from
Fig.~\ref{fig:abund} and Table~\ref{tab:abund} that the metallicity determined
from iron-peak elements is 0.09$\pm$0.09, and most elements have a relative
abundance close to this value. 

Lithium (Li/H=-10.38,
A(Li)=1.62) shows a considerable overabundance both relative to the solar
value (Li/H=-10.84, A(Li)=1.16) and relative to the normal red giants in the
spectra of which lithium is represented by a weak line or is not observed at all
\citep{2014ApJ...785...94L}. The absence of lithium is explained by its
depletion in the shell into which it is transferred by the red giant's
extensive convective envelope. The presence of lithium in the atmospheres of
some giants is associated with their activity \citep{1994ASPC...64..279F}, a
high rotation velocity \citep{1993ApJ...403..708F, 2013AN....334..120C}, and
even the presence of planets \citep{2014ApJ...785...94L}. Active \rs\ stars, to
which the star \pz\ being investigated belongs, generally exhibit a lithium
overabundance \citep{1993ChA&A..17...51L, 2003A&A...412..495M}. The derived
carbon (underabundance), nitrogen (overabundance), and oxygen (normal)
abundances are typical for normal red giants and reflect the result of the first
dredge-up \citep{1981ApJ...248..228L}. For the same reason, sodium also has a
slight overabundance, which is observed in red giants of various groups
\citep{2005ARep...49..535A, 2014AstL...40..406A}.

Figure~\ref{fig:abund} shows an overabundance of slow-neutron capture elements,
with the mean abundance of light s-elements (ls: Sr, Zr, Y, Mo) relative to iron
[ls/Fe] = 0.04$\pm$0.12 being appreciably smaller than the mean relative
abundance of heavy s-elements (hs: Ba, La, Ce, Nd) [hs/Fe] = 0.40$\pm$0.14. The
fact that the abundances of light s-elements were determined from the lines of
neutral atoms, for which the effect of departure from LTE reducing the the line
intensity is significant, while those of heavy ones were determined from ions in
the first dominant ionization stage, for which the effect of departure from LTE
may be neglected, can be responsible for such a difference. The spectral lines
of light s-elements in the second ionization stage are mostly in the blue part
of the spectrum, where the spectrum of \pz\ exhibits a strong blending due to
the influence of molecular lines and a high rotation velocity; therefore, we did
not use them. The effect of departure from LTE on the abundance being determined
can be significant. For example, it is $\sim$ 0.3 dex for zirconium
\citep{2010AstL...36..664V}. However, for the zirconium, strontium, and
molybdenum lines being used, the non-LTE corrections are difficult to estimate
due to the absence of such data. Thus, the observed difference in the abundances
of heavy and light s-elements [hs/ls] = 0.36 can be explained by the effect of
departure from LTE, while the overabundance of [s/Fe] itself can be, on average,
0.3--0.4~dex. An overabundance of s-elements was detected in the atmospheres of
classical barium and CH stars. However, they also exhibit a carbon
overabundance, in contrast to \pz. Moderate barium stars have a slight
overabundance of [s/Fe] at normal relative abundances of other elements
\citep{1977A&A....54..465P, 2002ARep...46..819B, 2007A&A...468..679S}. An
overabundance of s-elements is not typical for \rs\ stars.
\cite{2003A&A...412..495M} investigated the chemical composition of six active
\rs\ giants, and only one of them has the abundance [Ba/Fe] = 0.34.
\cite{2010OAP....23...17B} found no differences in the abundances of
neutron-capture elements in the atmospheres of more than 30 active (\rs\ and BY
Dra) and nonactive stars. Consequently, the moderate overabundance of s-elements
in the atmosphere of \pz\ is a characteristic of this star itself.

 The nature of moderate barium stars is not completely clear. These can be red
giants at a later evolutionary stage \citep{2002ARep...46..819B} and binary
stars with a relatively large separation of the components
\citep{1984ApJ...278..726B}. The binarity of some moderate barium stars was
discovered by the radial velocities (\citealp{1998A&AS..131...25U},
\citealp{1998A&AS..131...43U}); their periods ($P$) are thousands of days, while
the masses of the secondary components are close to the solar mass, which is
much larger than that for \pz\, but the ratios $M_2/\sqrt{P}$ are very close. In
addition, X-ray sources are encountered among the moderate barium stars
\citep{1996A&A...306..467J}, which may be indicative of their activity. Thus,
the nature of the overabundances of s-elements in the atmosphere of \pz\ may be
similar to the nature of those in moderate barium stars, but it is not typical
for all \rs\ stars and requires a more detailed study.

\section{Conclusions}

We determined the abundances of 26 chemical elements in the atmosphere of the
active red giant \pz\ by the method of model stellar atmosphere by taking into
account the hyperfine splitting, the isotopic shift, and (for some elements)
the departure from LTE through non-LTE corrections. Analysis of our data showed
that the abundances of most elements are typical for normal red giants, except
for lithium whose overabundance is characteristic for rapidly rotating and
active \rs\ stars and the neutron-capture elements whose overabundance may be
the same in nature as that in the atmospheres of moderate barium stars.

\section*{ACKNOWLEDGMENTS}

This work was supported in part by the Russian Foundation for Basic Research
(project no. 15-02-06046) and the Program of the Presidium of the Russian
Academy of Sciences ``Nonstationary Phenomena in the Universe''.

\bibliographystyle{mn2e}
\bibliography{paper}

\end{document}